\documentclass[submission,copyright,creativecommons]{eptcs}
\providecommand{\event}{SOS 2007} 

\usepackage{iftex}
\usepackage{amssymb}
\usepackage{amsthm}
\usepackage{amsmath}
\usepackage{bm}

\usepackage{tikz}
\usetikzlibrary{decorations.pathreplacing}

\newcommand{\Z}{{\mathbb Z}}
\newcommand{\area}{{\texttt{area}}}
\newcommand{\sper}{{\texttt{sper}}}
\newcommand{\xper}{{\texttt{xper}}}
\newcommand{\yper}{{\texttt{yper}}}
\newcommand{\Fib}{{\bm F}}

\ifpdf
  \usepackage{underscore}         
  \usepackage[T1]{fontenc}        
\else
  \usepackage{breakurl}           
\fi

\theoremstyle{plain}
\newtheorem{theorem}{Theorem}[section]

\theoremstyle{definition}

\title{Fibonacci and Catalan Numbers Meet in Staircase Polyominoes}

\author{Jean-Luc Baril
\institute{LIB, Université Bourgogne Europe,\\   Dijon, France}
\email{barjl@u-bourgogne.fr}
\and
Jos\'e L. Ram\'irez \qquad\qquad Samuel Ram\'irez
\institute{Departamento de Matem\'aticas, \\ Universidad Nacional de Colombia\\
Bogot\'a, Colombia}
\email{\quad jlramirezr@unal.edu.co \quad\qquad samramirezra@unal.edu.co}
\and 
Diego Villamizar \institute{Department of Mathematics, \\ Xavier University of Louisiana\\ New Orleans, LA 70125 \email{dvillami@xula.edu}}}
\def\titlerunning{Fibonacci and Catalan Numbers Meet in Staircase Polyominoes}
\def\authorrunning{J.-L. Baril \&  J. L. Ramírez \& S. Ramírez \& D. Villamizar}
\begin{document}
\maketitle

\begin{abstract}
We study Fibonacci (staircase) polyominoes, a class of column-convex polyominoes whose lower boundary is a staircase with unit vertical steps. We derive multivariate generating functions that refine Turban’s Fibonacci-number enumeration by tracking additional perimeter and area parameters. The proofs use a catalytic functional equation and, in a perimeter specialization, the kernel method, leading to explicit closed forms and Catalan-number coefficient formulas.
\end{abstract}

\section{Introduction}

The enumeration of polyomino classes defined by convexity and directionality constraints is a classical topic in enumerative combinatorics; see \cite{Book1} and the references therein. A \emph{polyomino} is a finite edge-connected union of unit squares in $\Z^2$. 

\begin{figure}[ht!]
\centering
\centering
\begin{tikzpicture}[scale=0.6, line cap=round, line join=round]
\tikzset{cell/.style={draw=black, very thick, fill=cyan!10!white}}

\foreach \x/\y in {0/0,0/1,0/2,0/3,0/4, 1/1, 1/2, 2/2, 2/3, 2/4, 2/5, 3/3, 3/4, 3/5, 3/6, 3/7,4/4, 4/5}{
  \draw[cell] (\x,\y) rectangle ++(1,1);
}

  \draw[very thick]
    (0,0) -- (1,0) -- (1,1) -- (2,1) --(2,2)--(3,2)--(3,3)--(4,3)--(4,4)--(5,4)--(5,5)--(6,5)--(6,6)--(7,6)--(7,7);
\end{tikzpicture}
\caption{Fibonacci polyomino of area 18.}\label{Ex1}
\end{figure}
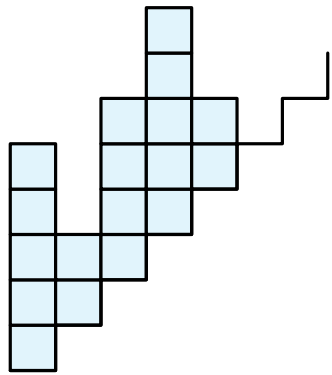

Among the many families studied with respect to parameters such as area and perimeter \cite{BleBreKnop3,Bou,ManSha2}, Turban \cite{Turban1} introduced a particularly natural class of \emph{Fibonacci polyominoes}. These are column-convex polyominoes whose lower boundary is a staircase path with unit vertical steps (see Figure~\ref{Ex1}). Turban proved that the number of such polyominoes of area $n$ equals the Fibonacci number $F_n$, and considered variants allowing staircase steps of arbitrary heights \cite{Turban1,Turban2}. 

In this paper we refine Turban’s enumeration by deriving multivariate generating functions that track additional parameters.

\section{Perimeter and area}

For a Fibonacci polyomino $P$, let $\area(P)$ denote its area (number of cells).
Write $\sper(P)$ for the semiperimeter, i.e., half of the number of unit edges
on the boundary of $P$. We further split it into the horizontal and vertical
contributions: $\xper(P)$ is half of the number of horizontal boundary edges (or number of columns of $P$) and
$\yper(P)$ is half of the number of vertical boundary edges. Thus
$\sper(P)=\xper(P)+\yper(P)$.

Let $\Fib$ be the class of all Fibonacci polyominoes. For $m,n,k\in\Z_{>0}$, let
$\Fib_{m,n,k}$ be the subset consisting of those $P$ with $2m$ horizontal boundary
edges, $2n$ vertical boundary edges, and area $k$, with $n\geq m$ and  $k\geq 2m-1$. Equivalently,
$\Fib=\bigcup_{m,n,k\ge 1}\Fib_{m,n,k}$. We encode these statistics by the
trivariate generating function
\[
F(x,y,q)=\sum_{P\in\Fib} x^{\xper(P)}\,y^{\yper(P)}\,q^{\area(P)}
=\sum_{m,n,k\ge 1}\left(\sum_{P\in\Fib_{m,n,k}}1\right)x^m y^n q^k.
\]

To keep track of the height of the first column, let $F_h(x,y,q)$ be the
generating function of Fibonacci polyominoes whose first column has height
exactly $h\ge1$, so that $F(x,y,q)=\sum_{h\ge1}F_h(x,y,q)$. Introducing a
catalytic variable $s$ marking this height, we set
\[
F(x,y,q;s)=\sum_{h\ge1} F_h(x,y,q)\,s^h,
\qquad\text{so that}\qquad F(x,y,q;1)=F(x,y,q).
\]
For brevity, write $F(s):=F(x,y,q;s)$ and $F(1):=F(x,y,q)$.

A decomposition according to the possible configurations of the final columns
yields the functional equation
\begin{equation}\label{eq:functionalareaper}
F(s)
= \frac{xyqs}{1-yqs} + \frac{xy (qs)^{2}}{1-qs}\, F(1)
+ \left( \frac{x (yqs)^{2}}{1-yqs} - \frac{xy(qs)^{2}}{1-qs} \right) F(sq).
\end{equation}
Iterating~\eqref{eq:functionalareaper} (with the convention that an empty product
equals $1$) and then setting $s=1$ gives an explicit closed form for $F(x,y,q)$.

\begin{theorem}\label{pol:areaper}
The generating function of nonempty Fibonacci polyominoes according to horizontal
semiperimeter, vertical semiperimeter, and area is
\[
F(x,y,q)
=\frac{
  \sum_{\ell\ge 0}
  \left(
    \frac{(xy)^{\ell+1} q^{(\ell+1)^{2}}(y-1)^{\ell}}{1-yq^{\ell+1}}
    \prod_{i=0}^{\ell-1}\frac{1}{(1-yq^{i+1})(1-q^{i+1})}
  \right)}{
  1-\sum_{\ell\ge 0}
  \left(
    \frac{(xy)^{\ell+1} q^{(\ell+1)(\ell+2)}(y-1)^{\ell}}{1-q^{\ell+1}}
    \prod_{i=0}^{\ell-1}\frac{1}{(1-yq^{i+1})(1-q^{i+1})}
  \right)}.
\]
\end{theorem}

As a consistency check, specializing~\eqref{eq:functionalareaper} at $s=x=y=1$
gives, with $F(q):=F(1,1,q)$,
\[
F(q)=\frac{q}{1-q}+\frac{q^2}{1-q}F(q),
\qquad\text{hence}\qquad
F(q)=\frac{q}{1-q-q^2}.
\]
Therefore the coefficient of $q^n$ is the Fibonacci number $F_n$, recovering
Turban’s enumeration by area.

\subsection{Perimeter}

We now specialize to perimeter enumeration by discarding the area statistic, that is, by setting $q=1$ in
\eqref{eq:functionalareaper}. In this specialization, the series $F(1)=F(x,y,1)$ is precisely the bivariate generating
function that counts Fibonacci polyominoes by horizontal and vertical semiperimeter.

Setting $q=1$ in \eqref{eq:functionalareaper} gives a linear functional equation in $F(s)$:
\[
\left(1-\frac{x(sy)^2}{1-sy}+\frac{xys^2}{1-s}\right)F(s)
=\frac{xys}{1-ys}+\frac{xys^2}{1-s}\,F(1).
\]

Let $p(m,n)$ be the number of Fibonacci polyominoes whose boundary has $2m$ horizontal edges and $2n$ vertical edges, and set
\[
P(x,y):=\sum_{m,n\ge1} p(m,n)\,x^m y^n.
\]
By definition, $P(x,y)=F(1)$. We compute it via the kernel method \cite{ban,pro}. Denote the kernel by
\[
K(s):=1-\frac{x(sy)^2}{1-sy}+\frac{xys^2}{1-s}.
\]
Choosing $s=s_0(x,y)$ so that $K(s_0)=0$ cancels the left-hand side and yields an explicit expression for $F(1)$ from the
right-hand side. We take the \emph{small} root (analytic at $x=0$), namely
\[
s_0=\frac{1+y+\sqrt{(1-y)(1-y-4xy)}}{2y\bigl(1+x(1-y)\bigr)}.
\]

\begin{theorem}\label{teorperimetro}
The generating function for nonempty Fibonacci polyominoes according to horizontal and vertical semiperimeter is
\[
P(x,y)=\frac{1 + x - y - 3xy -(1+x)\sqrt{(1-y)(1-y-4xy)}}{2x}.
\]
\end{theorem}

The first few terms of the expansion of $P(x,y)$ in powers of $y$ are
\begin{multline*}
 P(x,y)=   x y+\left(x^2+x\right) y^2+\left(2 x^3+3 x^2+x\right) y^3\\+\left(5 x^4+\bm{9 x^3}+5 x^2+x\right)
   y^4+\left(14 x^5+29 x^4+21 x^3+7 x^2+x\right) y^5+O\left(y^6\right).
\end{multline*}

\begin{figure}[ht!]
\centering
\begin{tikzpicture}[scale=0.5, line cap=round, line join=round]
\tikzset{cell/.style={draw=black, very thick, fill=cyan!10!white}}
\foreach \x/\y in {0/0,0/1,1/1,1/2,1/3,2/2}{
  \draw[cell] (\x,\y) rectangle ++(1,1);
}
\end{tikzpicture}
\hspace{0.6cm}
\begin{tikzpicture}[scale=0.5, line cap=round, line join=round]
\tikzset{cell/.style={draw=black, very thick, fill=cyan!10!white}}
\foreach \x/\y in {0/0,0/1,1/1,1/2,2/2,2/3}{
  \draw[cell] (\x,\y) rectangle ++(1,1);
}
\end{tikzpicture}
\hspace{0.6cm}
\begin{tikzpicture}[scale=0.5, line cap=round, line join=round]
\tikzset{cell/.style={draw=black, very thick, fill=cyan!10!white}}
\foreach \x/\y in {0/0,0/1,0/2,0/3,1/1,1/2,2/2}{
  \draw[cell] (\x,\y) rectangle ++(1,1);
}
\end{tikzpicture}
\hspace{0.6cm}
\begin{tikzpicture}[scale=0.5, line cap=round, line join=round]
\tikzset{cell/.style={draw=black, very thick, fill=cyan!10!white}}
\foreach \x/\y in {0/0,0/1,0/2,1/1,1/2,1/3,2/2}{
  \draw[cell] (\x,\y) rectangle ++(1,1);
}
\end{tikzpicture}
\hspace{0.6cm}
\begin{tikzpicture}[scale=0.5, line cap=round, line join=round]
\tikzset{cell/.style={draw=black, very thick, fill=cyan!10!white}}
\foreach \x/\y in {0/0,0/1,0/2,1/1,1/2,2/2,2/3}{
  \draw[cell] (\x,\y) rectangle ++(1,1);
}
\end{tikzpicture}
\vspace{0.5cm}

\begin{tikzpicture}[scale=0.5, line cap=round, line join=round]
\tikzset{cell/.style={draw=black, very thick, fill=cyan!10!white}}
\foreach \x/\y in {0/0,0/1,1/1,1/2,1/3, 2/2,2/3}{
  \draw[cell] (\x,\y) rectangle ++(1,1);
}
\end{tikzpicture}
\hspace{0.6cm}
\begin{tikzpicture}[scale=0.5, line cap=round, line join=round]
\tikzset{cell/.style={draw=black, very thick, fill=cyan!10!white}}
\foreach \x/\y in {0/0,0/1,0/2,0/3,1/1,1/2,1/3,2/2}{
  \draw[cell] (\x,\y) rectangle ++(1,1);
}
\end{tikzpicture}
\hspace{0.6cm}
\begin{tikzpicture}[scale=0.5, line cap=round, line join=round]
\tikzset{cell/.style={draw=black, very thick, fill=cyan!10!white}}
\foreach \x/\y in {0/0,0/1,0/2,1/1,1/2,1/3,2/2,2/3}{
  \draw[cell] (\x,\y) rectangle ++(1,1);
}
\end{tikzpicture}
\hspace{0.6cm}
\begin{tikzpicture}[scale=0.5, line cap=round, line join=round]
\tikzset{cell/.style={draw=black, very thick, fill=cyan!10!white}}
\foreach \x/\y in {0/0,0/1,0/2,0/3,1/1,1/2,1/3,2/2,2/3}{
  \draw[cell] (\x,\y) rectangle ++(1,1);
}
\end{tikzpicture}
\caption{Fibonacci polyominoes with 6 horizontal steps and 8 vertical steps.}\label{fig2}
\end{figure}
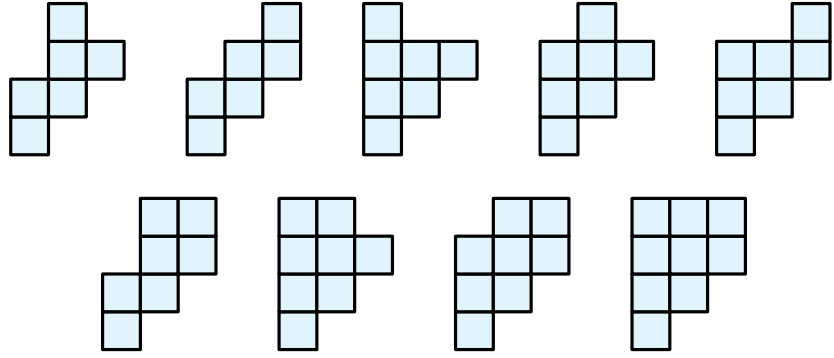

The bold coefficients in the above expansion correspond to the Fibonacci polyominoes shown in Figure~\ref{fig2}.

We denote by $C_n=\frac{1}{n+1}\binom{2n}{n}$ the $n$th Catalan number, and let 
$$C(z)=\sum_{n\geq 0}C_n z^n=\frac{1-\sqrt{1-4z}}{2z}$$
be its generating function.  From the previous theorem we obtain an explicit expression for the coefficients of $P(x,y)$ in terms of Catalan numbers. 

\begin{theorem}
 For $m\geq 1$ and  $n\geq 2$, we have 
 $$p(m,n)=[x^my^n]P(x,y)=C_m \binom{n - 2}{m - 1} + 
 C_{m - 1} \binom{n- 2}{m - 2}.$$
 In particular, for $n\geq 2$, $p(n,n)=C_{n-1}.$
  \end{theorem}

Notice that for the case $n=m$ we can give a combinatorial argument.  Let $P$ be a Fibonacci polyomino whose boundary has exactly $2n$ horizontal and $2n$ vertical edges. This implies that $P$ has $n$ columns. Observe that the bottom cell of each column produces (on its right side) a vertical  edge lying into the boundary. So, $P$ has exactly $n$ other vertical edges in the boundary, which implies that the number of cells in the $i$th column is at most $n-i+1$. Moreover, this also  implies that there is no descent in the polyomino, which means that the number of cells in the $i$th column is at most equal to the number plus one of cells in the $(i+1)$th column. Let $a_i$, $1\leq i\leq n$, be the number of cells in the column $i$. Then we have $2\leq a_i$ for $1\leq i\leq n-1$, $a_n=1$, and $a_i-2\leq a_{i+1}-1$, which proves that the word $b_1b_2\cdots b_{n-1}=(a_{n-1}-2)(a_{n-2}-2)\cdots (a_1-2)$ is a Catalan word, that is  $0\leq b_i\leq n-1$ and $b_{i+1}\leq b_{i}+1$. It is well known that these sequences are enumerated by the $(n-1)$th Catalan number (cf.\,\cite{CatWordsBargraphs}).

\begin{figure}[ht!]
\centering
\begin{tikzpicture}[scale=0.5, line cap=round, line join=round]
\tikzset{cell/.style={draw=black, very thick, fill=cyan!10!white}}
\foreach \x/\y in {0/0,0/1,1/1,1/2,2/2, 2/3, 3/3}{
  \draw[cell] (\x,\y) rectangle ++(1,1);
}
\end{tikzpicture}
\hspace{0.6cm}
\begin{tikzpicture}[scale=0.5, line cap=round, line join=round]
\tikzset{cell/.style={draw=black, very thick, fill=cyan!10!white}}
\foreach \x/\y in {0/0,0/1,0/2, 1/1,1/2,2/2,2/3, 3/3}{
  \draw[cell] (\x,\y) rectangle ++(1,1);
}
\end{tikzpicture}
\hspace{0.6cm}
\begin{tikzpicture}[scale=0.5, line cap=round, line join=round]
\tikzset{cell/.style={draw=black, very thick, fill=cyan!10!white}}
\foreach \x/\y in {0/0,0/1,1/1,1/2,1/3, 2/2, 2/3, 3/3}{
  \draw[cell] (\x,\y) rectangle ++(1,1);
}
\end{tikzpicture}
\hspace{0.6cm}
\begin{tikzpicture}[scale=0.5, line cap=round, line join=round]
\tikzset{cell/.style={draw=black, very thick, fill=cyan!10!white}}
\foreach \x/\y in {0/0,0/1,0/2,1/1,1/2,1/3,2/2, 2/3, 3/3}{
  \draw[cell] (\x,\y) rectangle ++(1,1);
}
\end{tikzpicture}
\hspace{0.6cm}
\begin{tikzpicture}[scale=0.5, line cap=round, line join=round]
\tikzset{cell/.style={draw=black, very thick, fill=cyan!10!white}}
\foreach \x/\y in {0/0,0/1,0/2,0/3, 1/1,1/2,1/3, 2/2,2/3, 3/3}{
  \draw[cell] (\x,\y) rectangle ++(1,1);
}
\end{tikzpicture}

\caption{Fibonacci polyominoes with 8 horizontal steps and 8 vertical steps.}\label{fig3}
\end{figure}
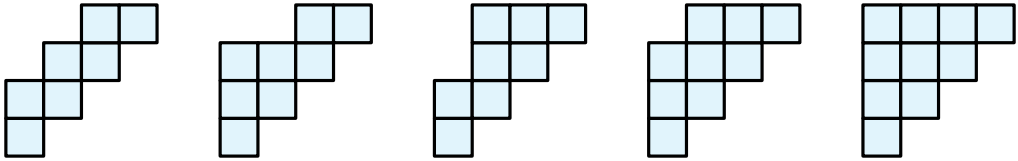

For example, the Catalan words of length $n=3$ are
$000, 001, 010, 011,$ and $012$. The corresponding Fibonacci polyominoes are shown in Figure~\ref{fig3}.

Notice that in the case $n = m$, $n$ is also the number of rows and the polyominoes are parallelogram polyominoes having the same number of rows and columns, so the upper boundary is an elevated Dyck path of length $2n$ and, consequently, they are enumerated by the $(n-1)$-th Catalan number.

\nocite{*}
\bibliographystyle{eptcs}
\bibliography{PaperGascom}

\begin{thebibliography}{1}
\providecommand{\bibitemdeclare}[2]{}
\providecommand{\surnamestart}{}
\providecommand{\surnameend}{}
\providecommand{\urlprefix}{Available at }
\providecommand{\url}[1]{\texttt{#1}}
\providecommand{\href}[2]{\texttt{#2}}
\providecommand{\urlalt}[2]{\href{#1}{#2}}
\providecommand{\doi}[1]{doi:\urlalt{https://doi.org/#1}{#1}}
\providecommand{\eprint}[1]{arXiv:\urlalt{https://arxiv.org/abs/#1}{#1}}
\providecommand{\bibinfo}[2]{#2}

\bibitemdeclare{article}{ban}
\bibitem{ban}
\bibinfo{author}{Cyril \surnamestart Banderier\surnameend},
  \bibinfo{author}{Mireille \surnamestart Bousquet-M{\'e}lou\surnameend},
  \bibinfo{author}{Alain \surnamestart Denise\surnameend},
  \bibinfo{author}{Philippe \surnamestart Flajolet\surnameend},
  \bibinfo{author}{Dani{\`e}le \surnamestart Gardy\surnameend} \&
  \bibinfo{author}{Dominique \surnamestart Gouyou-Beauchamps\surnameend}
  (\bibinfo{year}{2002}): \emph{\bibinfo{title}{Generating functions for
  generating trees}}.
\newblock {\slshape \bibinfo{journal}{Discrete Mathematics}}
  \bibinfo{volume}{246}(\bibinfo{number}{1--3}), pp. \bibinfo{pages}{29--55},
  \doi{10.1016/S0012-365X(01)00250-3}.

\bibitemdeclare{article}{BleBreKnop3}
\bibitem{BleBreKnop3}
\bibinfo{author}{Aubrey \surnamestart Blecher\surnameend},
  \bibinfo{author}{Charlotte \surnamestart Brennan\surnameend},
  \bibinfo{author}{Arnold \surnamestart Knopfmacher\surnameend} \&
  \bibinfo{author}{Toufik \surnamestart Mansour\surnameend}
  (\bibinfo{year}{2017}): \emph{\bibinfo{title}{The perimeter of words}}.
\newblock {\slshape \bibinfo{journal}{Discrete Mathematics}}
  \bibinfo{volume}{340}(\bibinfo{number}{10}), pp. \bibinfo{pages}{2456--2465},
  \doi{10.1016/j.disc.2017.06.003}.

\bibitemdeclare{article}{Bou}
\bibitem{Bou}
\bibinfo{author}{Mireille \surnamestart Bousquet-M{\'e}lou\surnameend} \&
  \bibinfo{author}{Andrew \surnamestart Rechnitzer\surnameend}
  (\bibinfo{year}{2003}): \emph{\bibinfo{title}{The site-perimeter of
  bargraphs}}.
\newblock {\slshape \bibinfo{journal}{Advances in Applied Mathematics}}
  \bibinfo{volume}{31}(\bibinfo{number}{1}), pp. \bibinfo{pages}{86--112},
  \doi{10.1016/S0196-8858(02)00553-5}.

\bibitemdeclare{article}{CatWordsBargraphs}
\bibitem{CatWordsBargraphs}
\bibinfo{author}{David \surnamestart Callan\surnameend},
  \bibinfo{author}{Toufik \surnamestart Mansour\surnameend} \&
  \bibinfo{author}{Jos{\'e}~L. \surnamestart Ram{\'i}rez\surnameend}
  (\bibinfo{year}{2021}): \emph{\bibinfo{title}{Statistics on Bargraphs of
  Catalan Words}}.
\newblock {\slshape \bibinfo{journal}{Journal of Automata, Languages and
  Combinatorics}} \bibinfo{volume}{26}(\bibinfo{number}{3--4}), pp.
  \bibinfo{pages}{177--196}, \doi{10.25596/jalc-2021-177}.

\bibitemdeclare{book}{Book1}
\bibitem{Book1}
\bibinfo{editor}{Anthony~J. \surnamestart Guttmann\surnameend}, editor
  (\bibinfo{year}{2009}): \emph{\bibinfo{title}{Polygons, Polyominoes and
  Polycubes}}.
\newblock {\slshape \bibinfo{series}{Lecture Notes in Physics}}
  \bibinfo{volume}{775}, \bibinfo{publisher}{Springer},
  \bibinfo{address}{Dordrecht}, \doi{10.1007/978-1-4020-9927-4}.

\bibitemdeclare{article}{ManSha2}
\bibitem{ManSha2}
\bibinfo{author}{Toufik \surnamestart Mansour\surnameend} \&
  \bibinfo{author}{Armend~Sh. \surnamestart Shabani\surnameend}
  (\bibinfo{year}{2019}): \emph{\bibinfo{title}{Enumerations on bargraphs}}.
\newblock {\slshape \bibinfo{journal}{Discrete Mathematics Letters}}
  \bibinfo{volume}{2}, pp. \bibinfo{pages}{65--94}.
\newblock \urlprefix\url{https://www.dmlett.com/archive/DML19_v2_p.65_94.pdf}.

\bibitemdeclare{article}{pro}
\bibitem{pro}
\bibinfo{author}{Helmut \surnamestart Prodinger\surnameend}
  (\bibinfo{year}{2004}): \emph{\bibinfo{title}{The kernel method: a collection
  of examples}}.
\newblock {\slshape \bibinfo{journal}{S{\'e}minaire Lotharingien de
  Combinatoire}} \bibinfo{volume}{50}, p. \bibinfo{pages}{B50f}.
\newblock
  \urlprefix\url{https://www.mat.univie.ac.at/~slc/wpapers/s50proding.html}.
\newblock \bibinfo{note}{19 pp.}

\bibitemdeclare{article}{Turban1}
\bibitem{Turban1}
\bibinfo{author}{Lo{\"i}c \surnamestart Turban\surnameend}
  (\bibinfo{year}{2000}): \emph{\bibinfo{title}{Lattice animals on a staircase
  and Fibonacci numbers}}.
\newblock {\slshape \bibinfo{journal}{Journal of Physics A: Mathematical and
  General}} \bibinfo{volume}{33}(\bibinfo{number}{13}), pp.
  \bibinfo{pages}{2587--2595}, \doi{10.1088/0305-4470/33/13/311}.

\bibitemdeclare{article}{Turban2}
\bibitem{Turban2}
\bibinfo{author}{Lo{\"i}c \surnamestart Turban\surnameend}
  (\bibinfo{year}{2006}): \emph{\bibinfo{title}{Lattice animals on a staircase
  and generalized Fibonacci numbers}}.
\newblock \doi{10.48550/arXiv.cond-mat/0106595}.
\newblock \eprint{cond-mat/0106595}.

\end{thebibliography}
\end{document}